\documentclass[final,1p,times,authoryear]{elsarticle}
\usepackage[plainpages=false, colorlinks=true, anchorcolor=blue, linkcolor=blue, citecolor=blue, bookmarks=false]{hyperref}
\usepackage[T1]{fontenc}
\usepackage{amsmath}
\usepackage{graphicx}
\usepackage{dcolumn}
\usepackage{bm}
\usepackage{adjustbox}
\usepackage{subcaption}
\usepackage{caption}
\newcommand{\rthis}[1]{\textcolor{black}{#1}}
\usepackage{float}

\begin{document}

\begin{frontmatter}
\journal{Journal of High Energy Astrophysics}

\title{Search for spatial coincidences between galaxy mergers and Fermi-LAT 4FGL-DR4 sources}

\author[1]{Siddhant Manna}
 \ead{Email:ph22resch11006@iith.ac.in}
\author[1]{Shantanu Desai}
 \ead{Email:shntn05@gmail.com}
\address[1]{
 Department of Physics, IIT Hyderabad Kandi, Telangana 502284,  India}





\begin{abstract}
We present a systematic search for spatial association between a high confidence sample of 3,166 morphologically selected galaxy mergers drawn from the initial convolutional neural network (CNN) catalog of 328,151 candidate mergers~\citep{Ackermann2018} and refined through SDSS cross-matching and the Fermi Large Area Telescope Fourth Source Catalog Data Release 4  (4FGL-DR4) catalog of gamma-ray sources detected by the Fermi Large Area Telescope (LAT). Using a conservative 4$\sigma$ positional uncertainty threshold and a Poisson based statistical framework, we identify 21 statistically significant associations with match probabilities $p < 0.05$. Among these are known classes of gamma-ray emitters such as Flat-Spectrum Radio Quasars (FSRQs), BL Lacertae objects (BLL), and radio galaxies bolstering the hypothesis that merger driven processes can fuel high energy activity. Intriguingly, five of the associated sources remain unclassified in 4FGL-DR4, hinting at a possible link between galaxy mergers and a hitherto unrecognized population of gamma-ray sources. The dominance of AGN like matches supports the scenario in which mergers trigger accretion onto central supermassive black holes, initiating AGN activity observable at gamma-ray energies. Moreover, the recurrent presence of unassociated sources among secure matches underscores the potential of merger catalogs as physically motivated priors in gamma-ray source identification efforts. This work constitutes the first dedicated effort to explore associations between gamma-ray sources and a large, morphologically selected sample of galaxy mergers~\citep{Ackermann2018} catalog, opening new avenues for understanding the role of galaxy interactions in high-energy astrophysics. We additionally examined the sample of 70 galaxy pairs from the Canadian Network for Observational Cosmology Field Galaxy Redshift Survey and found no statistically significant matches, with $p < 0.05$ within the 4$\sigma$ positional uncertainty threshold.
\end{abstract}

\end{frontmatter}
\section{\label{sec:level1}Introduction\protect}
The Fermi Large Area Telescope (LAT) has been a cornerstone in the exploration of the gamma-ray universe at GeV energies since its launch in 2008~\citep{Atwood2009}~\footnote{\url{https://fermi.gsfc.nasa.gov/science/instruments/lat.html}}. It has provided an unprecedented view of the high energy sky, enabling the detection and characterization of thousands of gamma-ray sources. The most recent data release, the Fourth Fermi-LAT Source Catalog Data Release 4 (4FGL-DR4) provides a comprehensive catalog of 7,195 gamma-ray sources observed over 14 years from 50 MeV to 1 TeV~\citep{Ballet2023} (see also the related 4FGL-DR3 catalog~\citealt{Abdollahi2022}). These sources include a plethora of astrophysical objects, such as blazars, radio galaxies, pulsars, etc.  However, a  large number of Fermi-LAT sources  ($\sim 2600$) remain unidentified and no counterparts at lower energies have been identified.
Numerous studies have therefore searched for spatial associations of Fermi 4FGL-DR4 sources with multi-wavelength source catalogs, including X-ray catalogs from SRG/eROSITA~\citep{Mayer2024} and SWIFT~\citep{Ulgiati2025}, radio source catalogs~\citep{Bruzewski2023}, South Pole Telescope (SPT) point sources at millimeter wavelengths~\citep{Zhang2022}, magnetars~\citep{Vyaas2024}, IceCube neutrino events~\citep{Negro2023}, MeerKAT sources~\citep{Himes2025} etc.

In this work, we search for statistically significant association between Fermi 4FGL-DR4 catalog and galaxy mergers. Galaxy mergers are pivotal events in the evolution of galaxies, driving significant morphological and dynamical changes that influence star formation, black hole growth, and active galactic nuclei (AGN) activity~\citep{Hopkins2006,Kaviraj2025}. These processes can produce energetic phenomena at gamma-ray energies, which can be detected by LAT. An important observational precedent for this scenario was reported by~\citet{Paliya2020}, who presented the first definitive case of a galaxy merger (TXS~2116$-$077) hosting a young, relativistic gamma-ray jet in a narrow line Seyfert~1 galaxy. 
This provides strong empirical support for the idea that merger triggered AGN activity can give rise to high energy gamma-ray emission~\citep{Paliya2020,Ramos2013,Chiaberge2015}.

Galaxy mergers are also thought to trigger intense star formation and fuel supermassive black holes, potentially leading to relativistic jets or other mechanisms responsible for gamma-ray production~\citep{Ellison2008,Ellison2011,Scudder2012,Satyapal2014,Gabor2016,Kashiyama2014}. Understanding the potential connection between galaxy mergers and gamma-ray sources is crucial for unraveling the physical processes governing these extreme astrophysical environments. A systematic search for associations between a comprehensive, high confidence catalog of galaxy mergers and the latest Fermi-LAT data release has not yet been done, to the best of our knowledge.

This work is heavily inspired by~\citet{Bouri2025}, which searched for a spatial association between IceCube-detected neutrinos and sources from six different galaxy merger catalogs, since neutrinos could be produced through shock acceleration from galaxy mergers, as discussed in various works~\citep{Lisenfeld2010,Kashiyama2014,Yuan2018}. 
Along with the associated neutrinos, one would also expect high energy gamma-ray radiation, which could be detected by Fermi-LAT. We briefly review some of the   gamma-ray production mechanisms discussed in literature. The first estimate of gamma-ray  flux from galaxy merger candidates  was done in ~\citet{Lisenfeld2010}. This work considered two specific merger systems, viz.  UGC 12914/5 and UGC 813/6~\citep{Lisenfeld2010}. These candidates are located at distances of 60 Mpc and 69 Mpc, respectively.   The hadronic gamma-ray flux  from $\pi^0$ decays resulting from energetic proton collisions with gas nuclei in the shocked region has been estimated in this fork to be $\mathcal{O} (10^{-15})~\rm{ph~cm^{-2}~s^{-1}}$, with a spectral index of 1.1 normalized at 1 TeV~\citep{Lisenfeld2010}. Although this flux is too small to be detected by Fermi-LAT, for a galaxy merger happening at distances between 2.6 and 3.9 Mpc, this flux could be up to 700 times higher, thereby enhancing the detection probability of  mergers in current generation gamma-ray telescopes such as Fermi-LAT. 

Then, ~\citet{Kashiyama2014} investigated the shock acceleration of particles in galaxy mergers with stellar masses of $\sim 10^{11} M_{\odot}$ and showed that cosmic rays could be accelerated up to 0.1-1 EeV with a spectral index of 2.0. It was shown that for a cosmic ray acceleration efficiency of 10\%,  one could detect gamma-ray radiation from about 10 nearby  merging galaxies with Cherenkov Telescope Array~\citep{Kashiyama2014}.  In addition to UGC 12914/5 and UGC 813/6, another galaxy merger candidate, viz. VV 114 located at a distance of $\sim 77$ Mpc~\citep{Vorontsov2001}  was identified as a potential gamma ray source. Subsequently,~\citet{Yuan2018} extended this analysis and calculated the cumulative diffuse gamma-ray background by incorporating mergers of galaxy cluster and group candidates, and also incorporated the redshift evolution of average galactic radius, shock velocity, and gas content inside haloes and galactic magnetic fields. This model is also consistent with the non-blazar contribution to extragalactic gamma-ray background~\citep{Yuan2018}. The empirical study of galaxy mergers and correctly identifying galaxies that have been merging has traditionally been quite challenging (See ~\citealt{Kaviraj2025} for a recent review).
Recent advances in catalogs from  large-scale optical surveys, such as the Sloan Digital Sky Survey Data Release 16 (SDSS DR16)~\citep{Ahmuda2020}, have however enabled the identification of large samples of galaxy mergers through morphological and spectroscopic signatures. 

The first catalog analyzed in this work is based on the convolutional neural network (CNN) catalog of 328,151 candidate mergers identified by~\citet{Ackermann2018} from which we filtered 3,166 galaxy mergers for our analysis; further details of which are provided in Section~\ref{sec:level2}. The  second catalog we considered for our analysis is the \textit{J/AJ/130/2043} catalog, from the Canadian Network for Observational Cosmology field galaxy redshift survey, contains 70 close galaxy pairs across four sky regions covering 1.5 square degrees, with redshifts in the range $0.1 \lesssim z \lesssim 0.6$~\citep{Patton2005}. We use both these aformentioned  catalogs for spatial coincidence with Fermi-LAT point sources.
These catalogs were also used for spatial coincidence analysis  with IceCube neutrinos  in~\citet{Bouri2025}.

We note that other major merger catalogs~\citep{Hwang2009,Gimeno2004} used in~\citet{Bouri2025} lack essential photometric or distance information required for flux-based probability estimates. Several well studied galaxy merger systems, such as UGC~12914/5, UGC~813/6, and VV~114, have been identified in prior literature as potential gamma ray sources due to their enhanced star formation and AGN activity triggered by gravitational interactions~\citep{Lisenfeld2010,Kashiyama2014}. Similarly, systems like NGC~660 and NGC~3256 are noted for their intense starburst and AGN driven processes, which could contribute to gamma-ray emission~\citep{Ackermann2018}. However, these systems are not included in the galaxy merger catalogs used in this study~\citep{Ackermann2018}. We searched for spatial associations between these systems and sources in the 4FGL-DR4 catalog 
within 1 arcminute radius, based on the positional accuracy of the catalog, but found no matches. A targeted analysis of these individual systems is being conducted in a follow-up study to further explore their gamma-ray emission potential. 

While earlier studies have explored potential links between galaxy mergers and high-energy neutrino emission~\citep{Bouri2025}, no systematic effort has yet been made to investigate such connections in the gamma-ray regime. Our work represents the first dedicated search for associations between gamma-ray sources and a large, morphologically selected sample of galaxy mergers~\citep{Ackermann2018}, thereby opening new avenues for probing the role of galaxy interactions in shaping high-energy astrophysical phenomena. 

This paper is organized as follows. Section~\ref{sec:da} describes the data selection and cross-matching methodology, Section~\ref{sec:results} presents the findings, section~\ref{sec:conclusion} discusses the conclusions from our present study, and finally section~\ref{sec:Discussion and Outlook} summarizes implications and possible follow-up directions.

\section{Data Analysis}
\label{sec:da}
\label{sec:level2}
We analyze the aformentioned galaxy merger catalog that contains 328,151 candidate mergers~\citep{Ackermann2018}.  They classified galaxy mergers in SDSS imaging data using deep convolutional neural networks (CNNs)~\citep{Krizhevsky2017}, specifically adopting the {\tt Xception} architecture~\citep{Chollet2016}. Their approach employed transfer learning by initializing network weights from models pre-trained on the ImageNet dataset~\citep{Deng2009}, which is a huge dataset of a few
million natural images from thousands of categories, like cats, dogs and many others. The classifier was trained on a labeled dataset derived from SDSS Data Release 7~\citep{Abazajian2009}, consisting of RGB JPEG cutouts obtained via the SDSS image cutout service~\footnote{\url{https://skyserver.sdss.org/dr12/en/help/docs/api.aspx}}. Ground truth labels were sourced from the Galaxy Zoo project~\citep{Lintott2008,Lintott2011}, which offers crowd sourced morphological classifications based on visual inspection by citizen scientists. The positive class comprised 3,003 visually confirmed mergers from the catalog of~\citet{Darg2010}, selected using a weighted merger vote fraction $( f_m > 0.4 )$ and limited to the redshift range $( 0.005 < z < 0.1 )$. The negative class included 10,000 randomly selected non interacting galaxies from the Galaxy Zoo project for the same redshift interval, with $( f_m < 0.2 )$, serving as a control sample.
To mitigate class imbalance during training, stratified sampling was employed to construct balanced mini batches with equal numbers of merger and non-merger galaxies. Model performance was evaluated using 4-fold cross-validation, with class distributions preserved across folds. Training was carried out on subsets of varying sizes (3,000, 1,500, 900, 500, and 300 samples), demonstrating that transfer learning consistently improved performance.

\begin{figure}[ht]
\centering
\includegraphics[width=\linewidth]{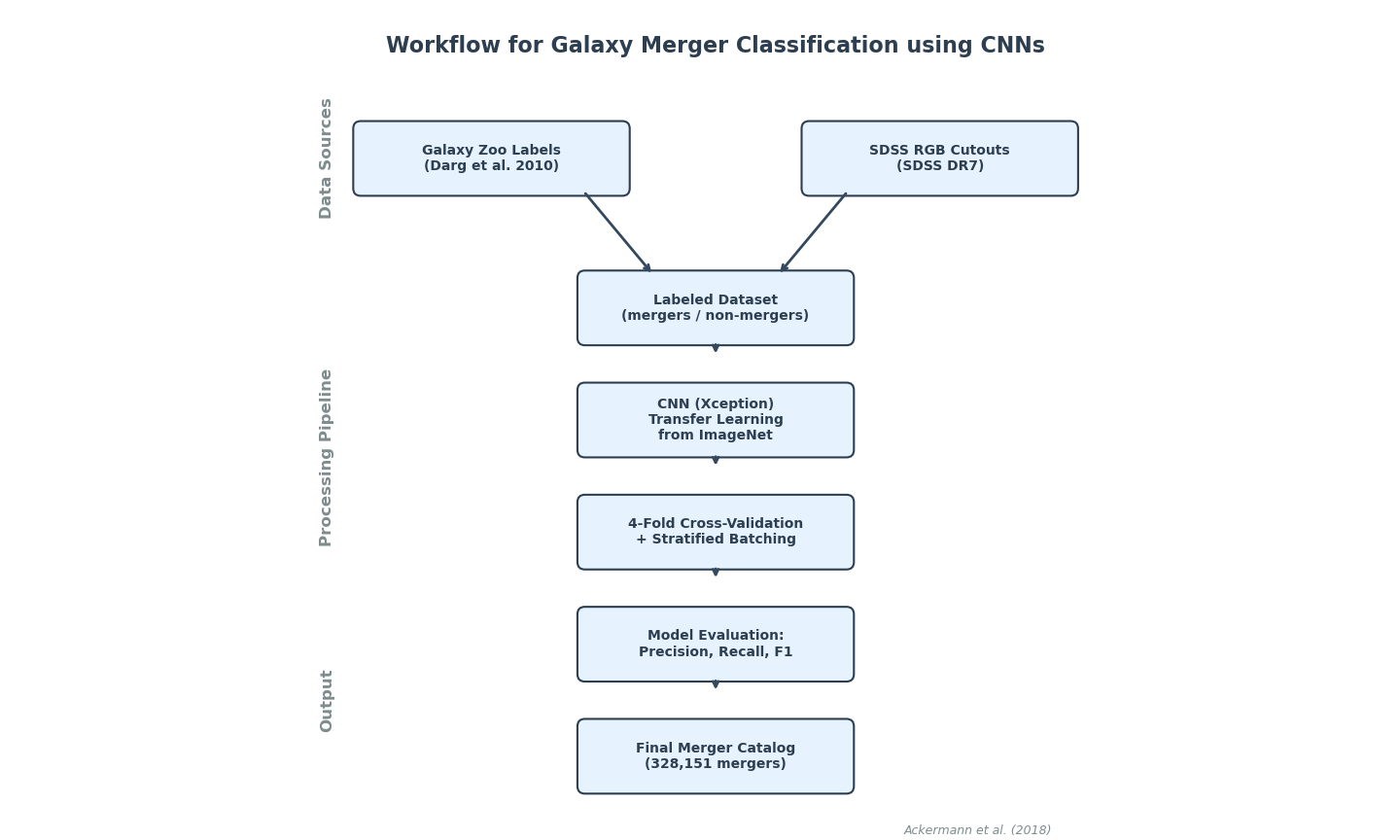}
\caption{Methodological workflow used in \citet{Ackermann2018}, illustrating the training and evaluation pipeline for galaxy merger classification using CNNs and transfer learning from ImageNet. This figure has been adapted from ~\citet{Ackermann2018}.}
\label{fig:Figure1}
\end{figure}

Following training, the CNN was evaluated on a separate test set that was never used during training. For each test galaxy, the model produced a continuous score$( p_m \in [0,1] )$\, representing the predicted probability of the galaxy being a merger. A threshold of $( p_m = 0.5 )$ was used to assign binary labels, galaxies with $( p_m > 0.5 )$ were classified as mergers, and those with $( p_m \leq 0.5 )$ as non-interacting systems. Classification performance was assessed using standard evaluation metrics, including precision, recall, and F1-score.
Their method achieved a precision of 0.97, recall of 0.96, and F1-score of 0.97, surpassing the performance of earlier merger classification techniques such as those proposed by earlier works~\citep{Goulding2018,Cotini2013,Hoyos2012}. The results highlight the high reliability and robustness of the CNN-based approach in identifying galaxy mergers from imaging data.
While the CNN classifier demonstrates strong performance in merger identification, several factors may contribute to biases in its classification accuracy. First, the training labels are derived from Galaxy Zoo, a citizen science project involving human visual classification. These labels, while statistically robust in aggregate, are subject to individual perceptual biases and inconsistencies, which may be implicitly learned by the CNN. Second, class definitions are based on thresholds of the merger vote fraction ($f_m$), with mergers defined by $f_m > 0.4$ and non-mergers by $f_m < 0.2$. Galaxies with intermediate values are excluded, potentially introducing boundary effects and under representing ambiguous cases. Third, transfer learning from the ImageNet domain may introduce domain mismatch. In Figure~\ref{fig:Figure1} we showcase the pictorial representation of the workflow followed by~\citet{Ackermann2018}.

\begin{figure}[ht]
\centering
\includegraphics[width=\linewidth]{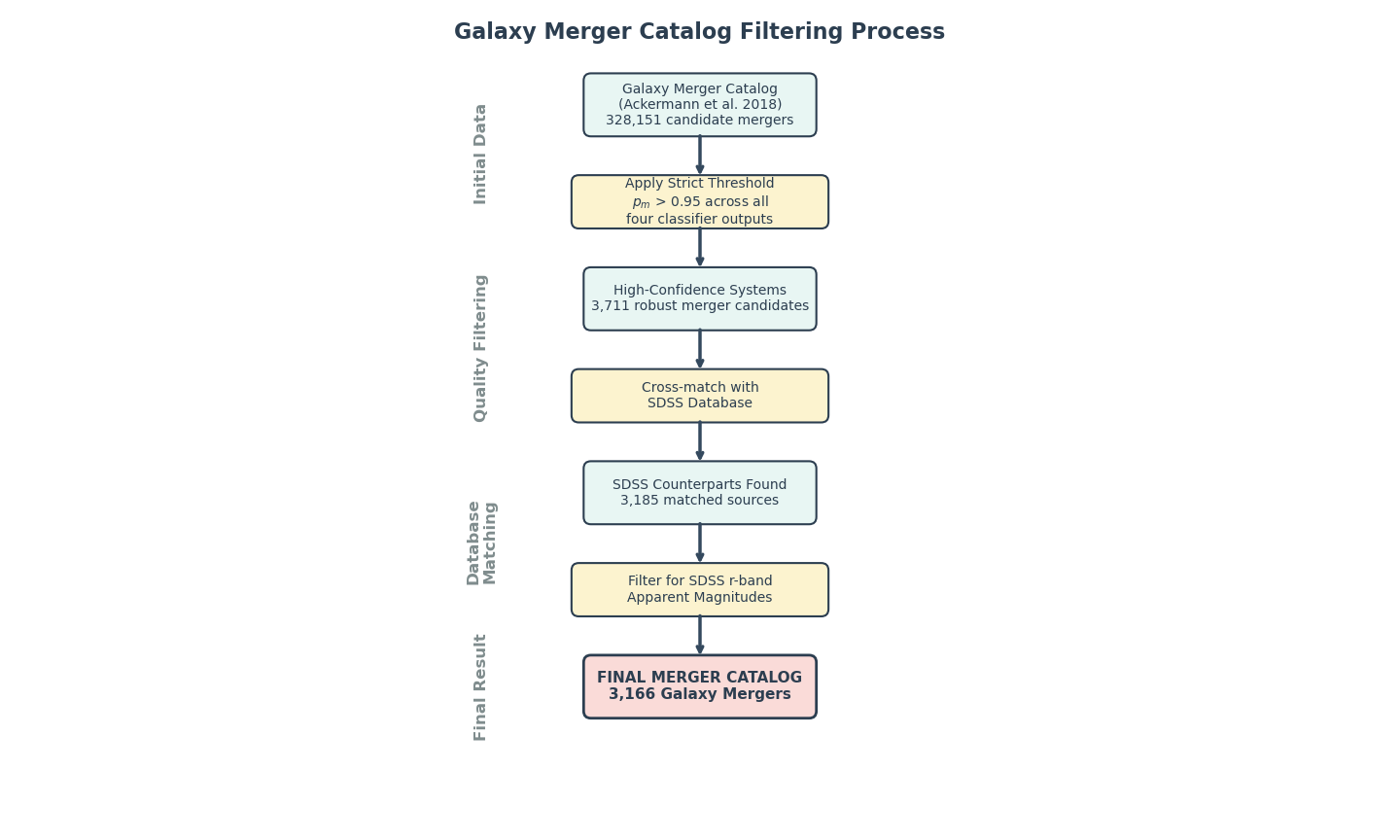}
\caption{Filtering workflow for galaxy merger catalog refinement. The process reduces 328,151 initial candidates ~\citet{Ackermann2018} to 3,166 high-confidence mergers through strict probability thresholds ($p_m > 0.95$), SDSS cross-matching, and photometric quality cuts.}
\label{fig:Figure2}
\end{figure}

To isolate high-confidence systems, we applied a strict threshold of $p_m > 0.95$ across all four classifier outputs, retaining only those galaxies that satisfied this criterion in every fold. This conservative approach yielded a subset of 3,711 robust merger candidates. This stricter criterion minimizes the inclusion of ambiguous cases and emphasizes classifier agreement, albeit at the cost of reducing the final sample size. These selected objects were cross matched with the Sloan Digital Sky Survey Data Release 16 (SDSS DR16) catalog~\citep{Ahmuda2020} using TOPCAT (Tool for OPerations on Catalogues And Tables, version 4.10-3), an interactive software tool for manipulating and visualizing large astronomical datasets~\citep{Taylor2005}. TOPCAT facilitates efficient data handling through functionalities like filtering, cross-matching, and plotting, making it ideal for processing astrophysical survey data. A matching radius of 0.1 arcseconds was adopted to ensure precise associations.
The resulting cross-match identified 3,185 mergers with SDSS counterparts. Among these matched sources, 1,147 have measured spectroscopic redshifts and 3,166 sources possess SDSS $r$-band apparent magnitudes. We show our analysis workflow in Figure~\ref{fig:Figure2}.

\begin{figure}[ht]
    \centering
    \includegraphics[width=0.9\textwidth]{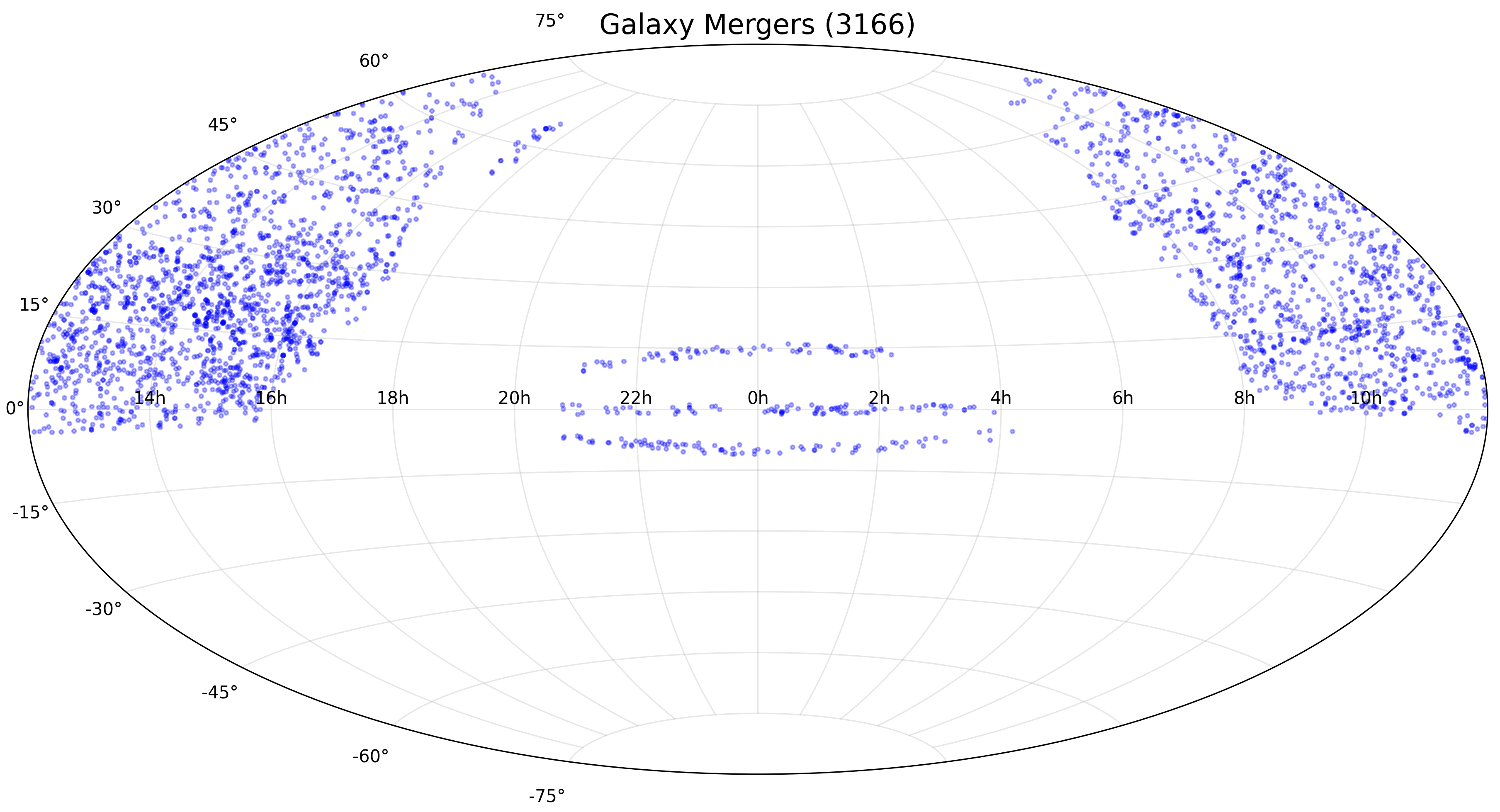}
    \caption{Aitoff projection showing the sky distribution of 3,166 galaxy mergers~\citep{Ackermann2018} with SDSS $r$-band magnitudes. Positions are plotted in equatorial coordinates (RA, Dec).}
    \label{fig:Figure3}
\end{figure}

To characterize the spatial extent of the merger sample, we estimated the total angular sky area over which the 3,166 galaxy mergers are distributed. This effective footprint was computed by dividing the sky into finer bins in right ascension and declination and summing up the area of all the spatial bins that contained at least one merger. The contribution of each bin was corrected for spherical projection by weighting with cos($\delta$), where $\delta$ is the central declination of the bin. This produced an estimated sky coverage of approximately \(6190~\mathrm{deg}^2\), which was used for surface density estimates and association probability calculations. Figure~\ref{fig:Figure3} shows the sky distribution in equatorial coordinates using an Aitoff projection.
For the gamma-ray dataset, we utilized 4FGL-DR4 catalog, which contains 7,195 gamma-ray sources detected between 50 MeV and 1 TeV over 14 years of observations~\citep{Ballet2023}.

To identify possible associations, we searched for candidate galaxy mergers located within the 4$\sigma$ positional uncertainty ellipse of each 4FGL-DR4 source. This conservative threshold, similar to that used in~\citet{Zhang2022}, accounts for both Fermi-LAT's positional uncertainties and uncertainties in the external catalogs. When multiple optical sources were found within a single gamma-ray error ellipse, we assessed the probability of random association using a Poisson-based framework as described below~\citep{Browne1978, Downes1986, Biggs2011, Zhang2022}.

The expected number of chance associations $(\mu$) can be calculated as follow:~\citep{Browne1978, Downes1986, Biggs2011,Zhang2022}:
\begin{equation}
    \mu = \pi a b \, n_s(>S),
    \label{eq:mu}
\end{equation}
\noindent where $a$ and $b$ are the semi-major and semi-minor axes of the Fermi positional ellipse, and $n_s(>S)$ is the surface density of mergers with flux greater than that of the SDSS counterpart having flux equal to $S$. To evaluate $n_s(>S)$, we used the SDSS-derived fluxes computed from the $r$-band magnitudes of the merger candidates. For each candidate to be tested, we determined the number of mergers in the full sample with $r$-band fluxes exceeding the corresponding candidate flux value. The details of the  calculation of $r$-band flux from the estimated magnitude can be found in  ~\ref{appendix:sdss_flux_conversion}. 
Dividing this count by the total sky area covered by the catalog yielded a flux-dependent surface density. This value of $n_s(>S)$ was then used in Eq.~\ref{eq:mu} to calculate the expected number of chance associations ($\mu$), allowing us to assess the probability of random coincidence.

The probability of random coincidence (i.e., the \textit{p}-value) is then given by:
\begin{equation}
    p = 1 - e^{-\mu}.
    \label{eq:pvalue}
\end{equation}
A small value $p$ indicates a low probability of random association, which implies a more reliable counterpart match. We adopt a primary threshold of $p < 0.05$ to define a secure association, and a secondary, more relaxed threshold of $p < 0.1$ to capture additional statistically plausible matches, following the approach of~\citep{Zhang2022}. In this work, the authors conducted a systematic cross matching of Fermi-LAT 4FGL sources with multi wavelength catalogs (radio, millimeter, and X-ray) in the South Pole Telescope Sunyaev-Zel'dovich  (SPT-SZ) survey region. They adopted a matching radius of 4$\sigma$ positional uncertainty and a $p$-value threshold of $p < 0.1$ to identify statistically plausible associations. They demonstrated that while the probability threshold of $p < 0.05$ ensures high purity, it may exclude real counterparts, especially in crowded fields or when positional uncertainties are large. By extending the threshold to $p < 0.1$, they achieved a better trade off between completeness and reliability, which maintained low contamination levels under the relaxed criterion.

Given the rarity and low surface density of galaxy mergers, we expect the false match rate to remain low even with a relaxed threshold~\citep{Downes1986, Biggs2011}. 

\subsection{Data Provenance}
The galaxy merger catalog and associated convolutional neural network (CNN) classifications are adopted directly from \citet{Ackermann2018}. 
These include the merger probability scores derived using the {\tt Xception} architecture and their validation against Galaxy Zoo labels. 
\rthis{The contributions of this work consist of: }(i) applying a stricter probability threshold ($p_m > 0.95$) to define a robust subset of mergers, 
(ii) cross-matching these mergers with SDSS DR16 counterparts, (iii) calculating flux-dependent surface densities and sky area coverage, 
(iv) statistical validation of merger--gamma-ray associations using a Poisson-based framework, and 
(v) control-sample analysis to assess chance coincidence rates. 
This delineation clarifies the division between previously published data products and the new analyses carried out in this study.

\section{Results}
\label{sec:results}
We began our analysis by examining a supplementary sample of 70 galaxy pairs from the Canadian Network for Observational Cosmology Field Galaxy Redshift Survey, as defined in~\citet{Patton2005}. Applying the same matching criterion and statistical methodology, we found no statistically significant associations with Fermi 4FGL-DR4 sources within the 4$\sigma$ positional uncertainty threshold $p < 0.05$.  Therefore, we do not find any Fermi 4GL-DR4 counterparts in  the  \textit{J/AJ/130/2043} merger catalog.

Next, we applied the matching criterion defined in Equation~\ref{eq:pvalue} to our primary sample of 3,166 morphologically selected galaxy mergers identified from SDSS imaging. This resulted in identification of 38 spatial coincidences between Fermi 4FGL-DR4 sources and galaxy mergers from our sample.
To assess the likelihood of true physical associations versus chance alignments, we computed Poisson based match probabilities for each pair using Eq.~\ref{eq:pvalue}. Among these 38 matches, 21 exhibit \emph{p}-values below 0.05, indicating a low probability of random association. We classify these as statistically secure matches. 
According to source classifications in the 4FGL-DR4 catalog, the 21 matches within the 4$\sigma$ ellipses include four Flat Spectrum Radio Quasars (FSRQs), six BL Lacertae objects (BLL), three Blazar Candidates of Uncertain type (BCUs), and three Radio Galaxies (RDGs). The remaining five are unassociated sources (UIDs) lacking confirmed multiwavelength counterparts. Notably, none of the secure matches is classified as a pulsar. The five UIDs span gamma-ray flux densities in the range $(0.55\text{--}1.90) \times 10^{-35}\,\mathrm{erg\,cm^{-2}\,s^{-1}\,Hz^{-1}}$, suggesting the possible existence of a previously unrecognized population of gamma-ray sources associated with galaxy mergers. 

\begin{figure}[htbp]
    \centering
    \includegraphics[width=0.8\textwidth]{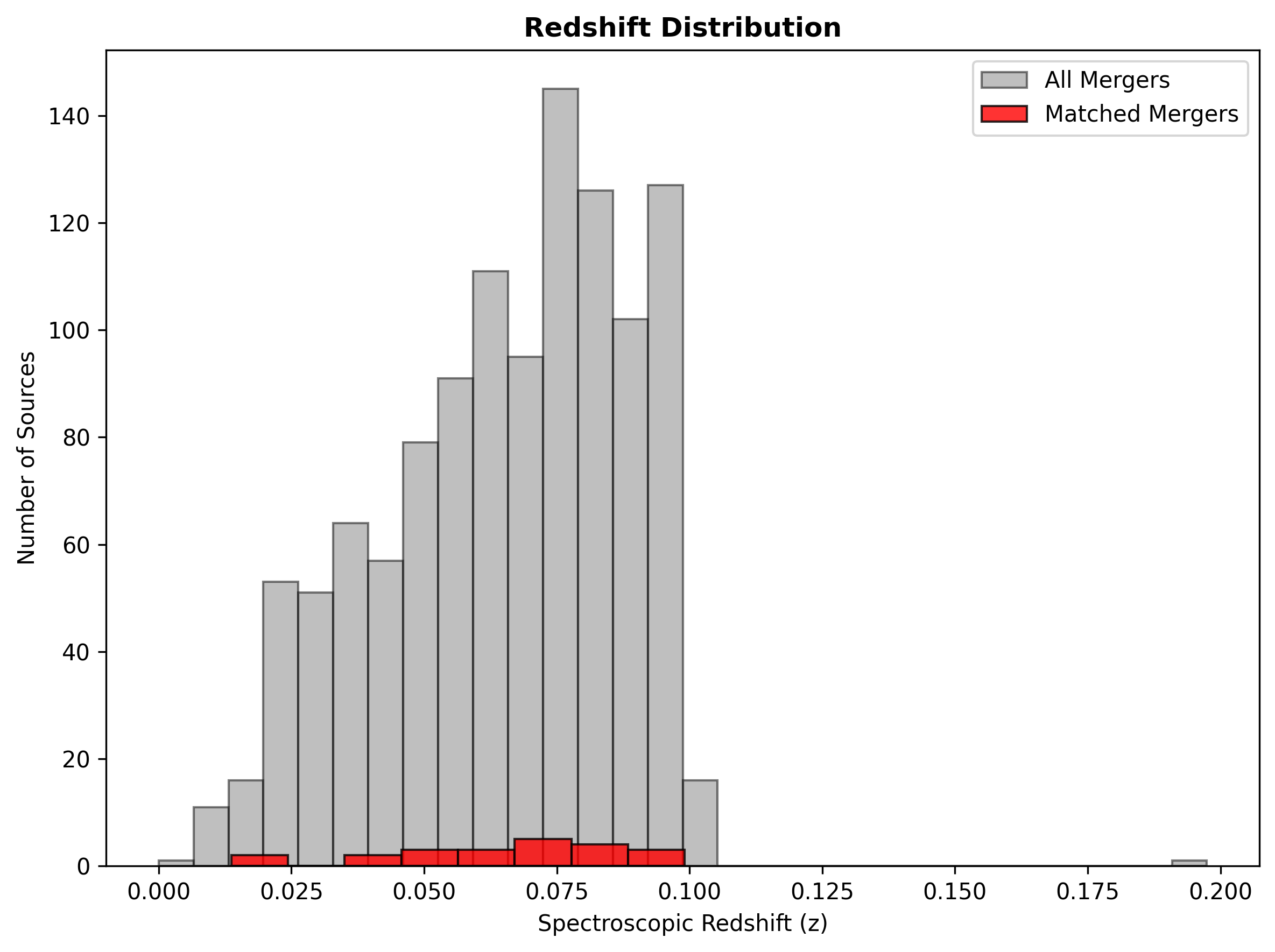}
    \caption{Redshift distribution comparison between the full merger catalog (gray) and the 21 matched mergers (red). The matched sample shows a similar redshift distribution to the overall population, with most sources at $z < 0.1$.}
    \label{fig:Figure4}
\end{figure}

\begin{figure}[htbp]
    \centering
    \includegraphics[width=0.8\textwidth]{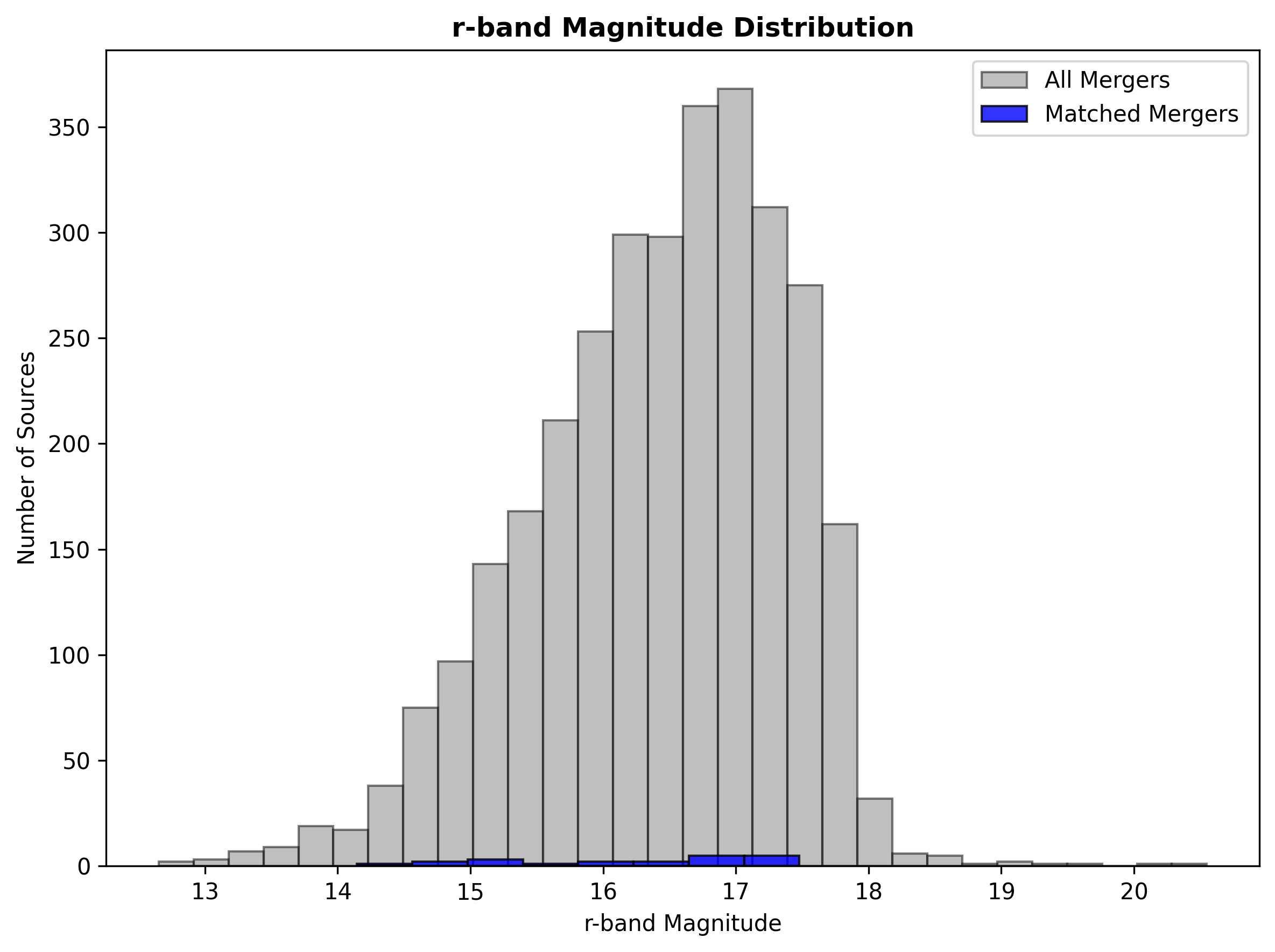}
    \caption{$r$-band magnitude distribution comparison between the full merger catalog (gray) and the 21 matched mergers (blue). The matched sample spans a range of magnitudes from $\sim$14 to $\sim$17.5 mag, representative of the brighter portion of the full sample.}
    \label{fig:Figure5}
\end{figure}

\begin{figure}[htbp]
    \centering
    \includegraphics[width=0.7\textwidth]{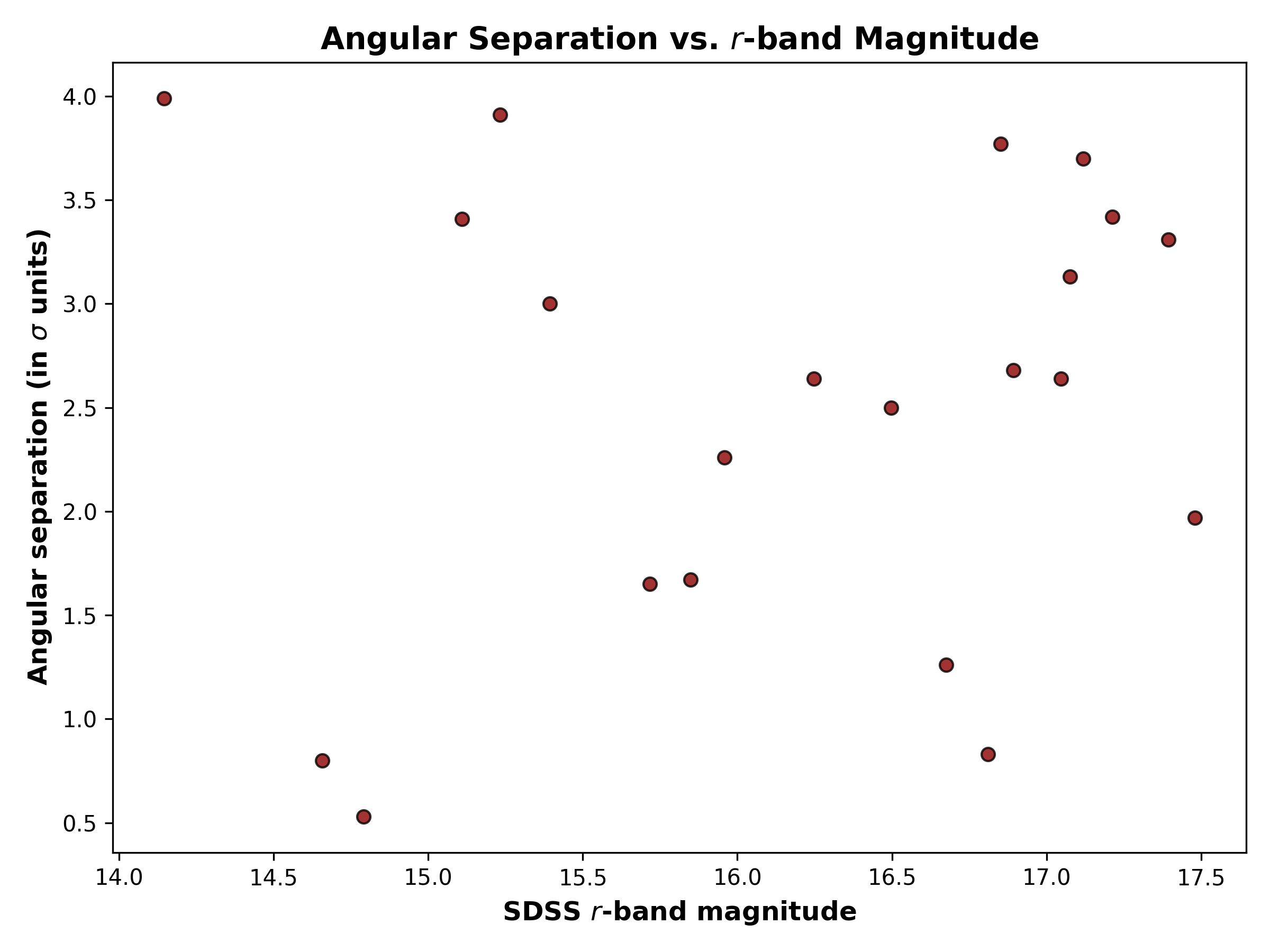}
    \caption{Angular separation (in units of positional uncertainty $\sigma$) between matched Fermi 4FGL-DR4 sources and their associated SDSS galaxy mergers, plotted against the $r$-band magnitude. This helps evaluate whether brighter mergers yield more spatially consistent associations compared to fainter counterparts.}
    \label{fig:Figure6}
\end{figure}

To investigate potential observational selection effects, we also compared the distributions of spectroscopic redshift ($z$) and $r$-band magnitude for the full merger catalog and the subset of 21 statistically significant matched sources ($p < 0.05$). As shown in Figure~\ref{fig:Figure4}, the redshift distribution of the matched mergers closely follows that of the entire catalog, with the majority of sources lying at $z < 0.1$. This indicates that the associations are not strongly biased toward unusually nearby systems.
Next, we compared the $r$-band magnitude distribution of the full merger sample consisting of 3,166 galaxies with that of the 21 matched mergers, as shown in Figure~\ref{fig:Figure5}. The matched mergers are found to span magnitudes in the range $\sim14$--$17.5$, corresponding to the brighter portion of the parent distribution.
We also did additional tests to explore whether the quality of spatial associations depends on the optical brightness of the merger counterparts, we plot the angular separation, expressed in units of the Fermi source's 95\% positional uncertainty ($\sigma$) against the SDSS $r$-band magnitude of the matched mergers in Figure~\ref{fig:Figure6}. We find no evidence for a correlation, indicating that the spatial consistency of the associations does not depend on the optical brightness of the mergers.

\begin{figure*}[htbp]
    \centering
    \renewcommand{\thesubfigure}{}

    \begin{subfigure}[b]{0.18\textwidth}
        \includegraphics[width=\textwidth]{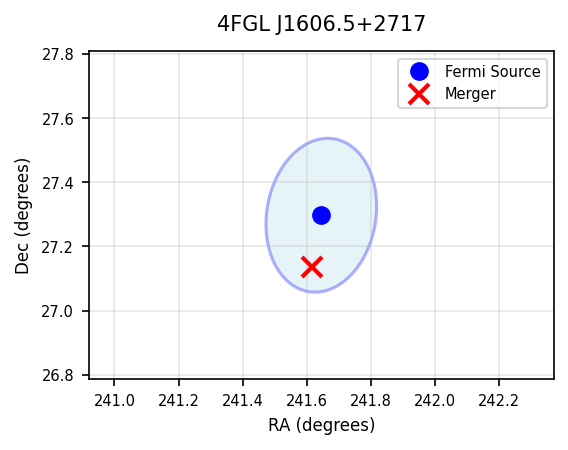}
        \caption*{Source 1}
    \end{subfigure}
    \begin{subfigure}[b]{0.18\textwidth}
        \includegraphics[width=\textwidth]{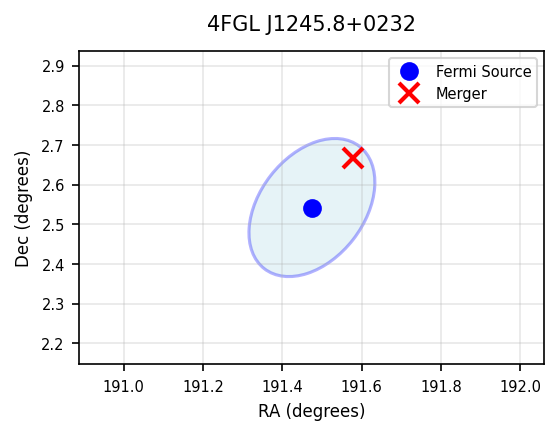}
        \caption*{Source 2}
    \end{subfigure}
    \begin{subfigure}[b]{0.18\textwidth}
        \includegraphics[width=\textwidth]{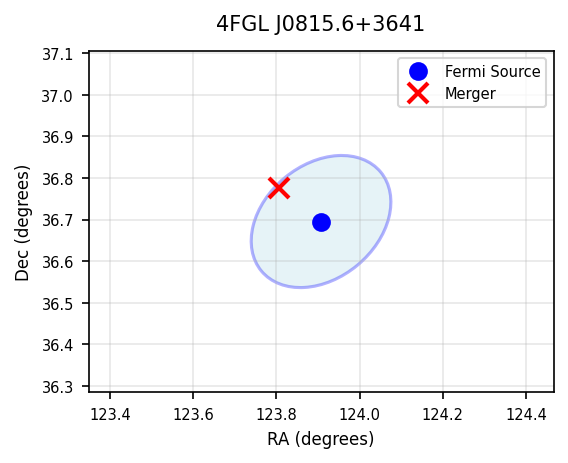}
        \caption*{Source 3}
    \end{subfigure}
    \begin{subfigure}[b]{0.18\textwidth}
        \includegraphics[width=\textwidth]{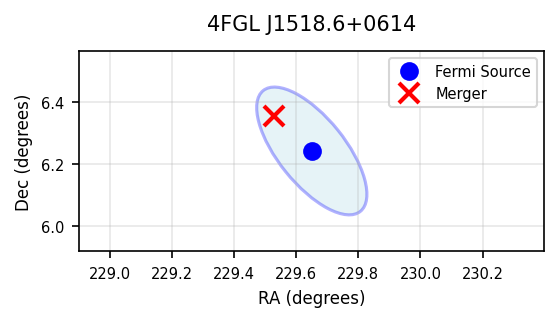}
        \caption*{Source 4}
    \end{subfigure}
    \begin{subfigure}[b]{0.18\textwidth}
        \includegraphics[width=\textwidth]{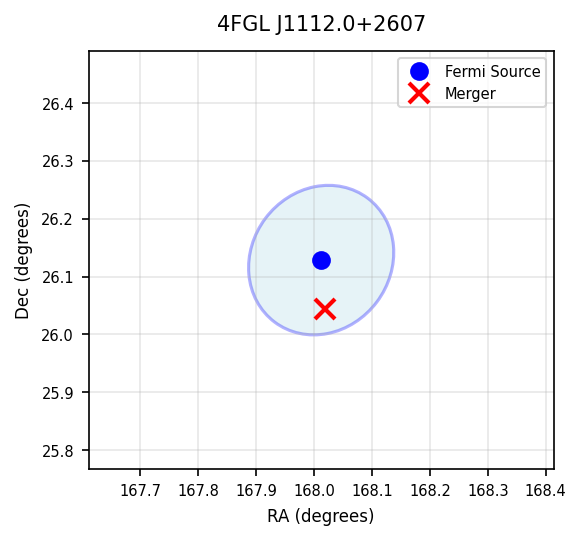}
        \caption*{Source 5}
    \end{subfigure}

    \vspace{0.2cm}

    \begin{subfigure}[b]{0.18\textwidth}
        \includegraphics[width=\textwidth]{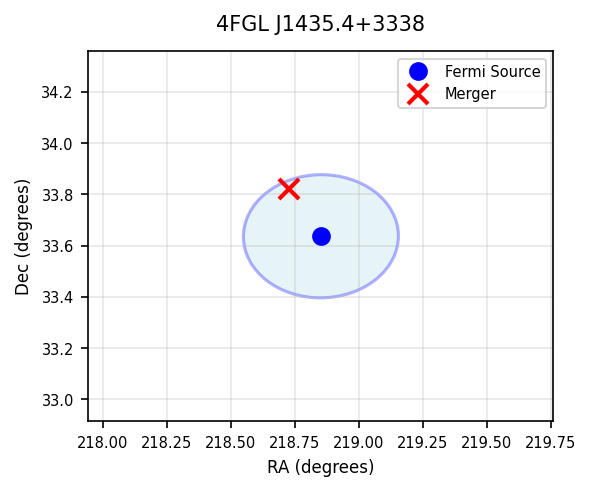}
        \caption*{Source 6}
    \end{subfigure}
    \begin{subfigure}[b]{0.18\textwidth}
        \includegraphics[width=\textwidth]{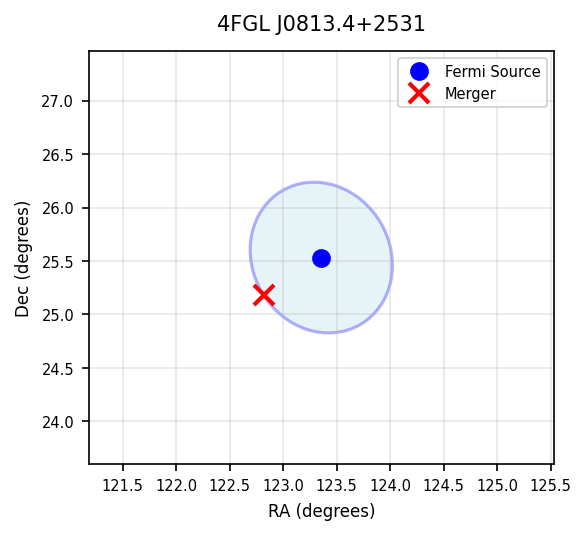}
        \caption*{Source 7}
    \end{subfigure}
    \begin{subfigure}[b]{0.18\textwidth}
        \includegraphics[width=\textwidth]{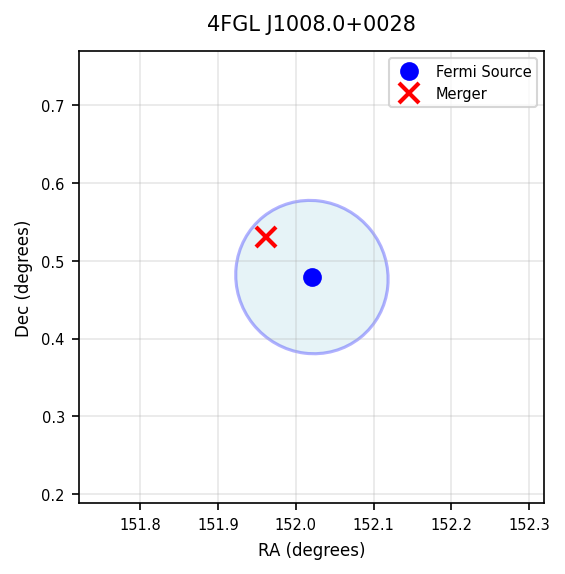}
        \caption*{Source 8}
    \end{subfigure}
    \begin{subfigure}[b]{0.18\textwidth}
        \includegraphics[width=\textwidth]{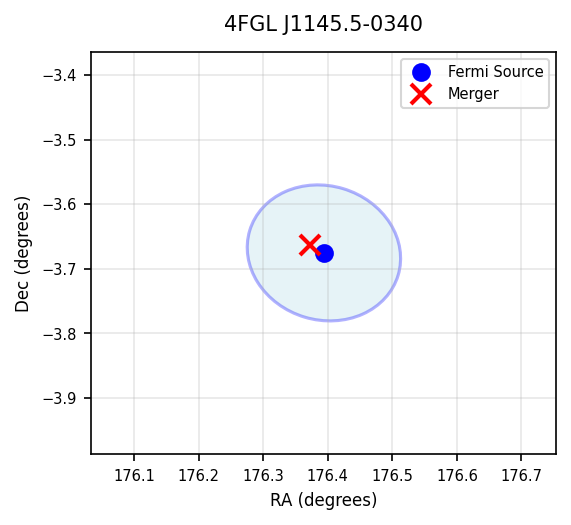}
        \caption*{Source 9}
    \end{subfigure}
    \begin{subfigure}[b]{0.18\textwidth}
        \includegraphics[width=\textwidth]{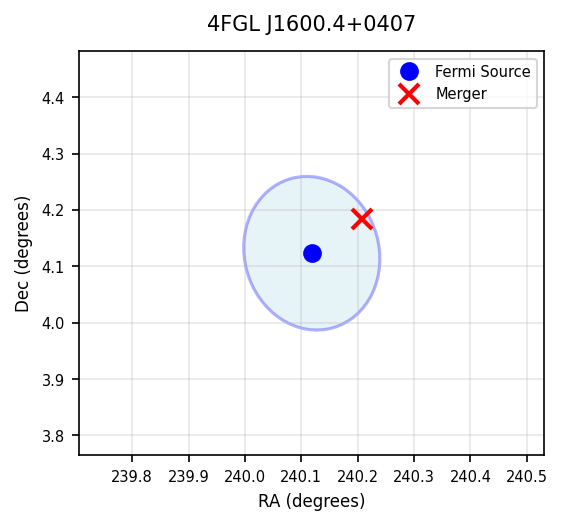}
        \caption*{Source 10}
    \end{subfigure}

    \vspace{0.2cm}

    \begin{subfigure}[b]{0.18\textwidth}
        \includegraphics[width=\textwidth]{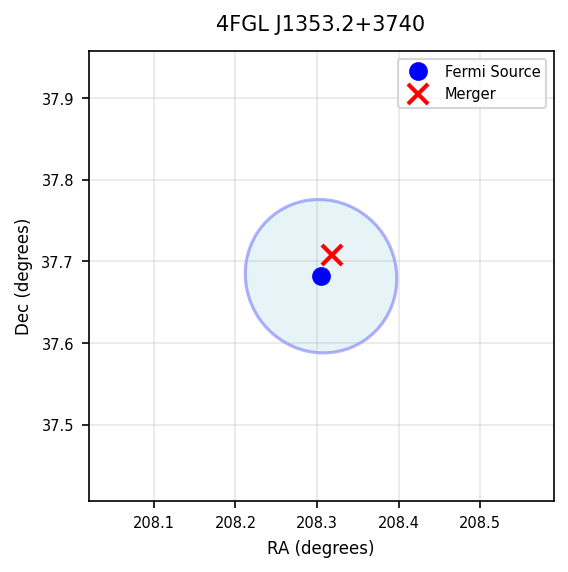}
        \caption*{Source 11}
    \end{subfigure}
    \begin{subfigure}[b]{0.18\textwidth}
        \includegraphics[width=\textwidth]{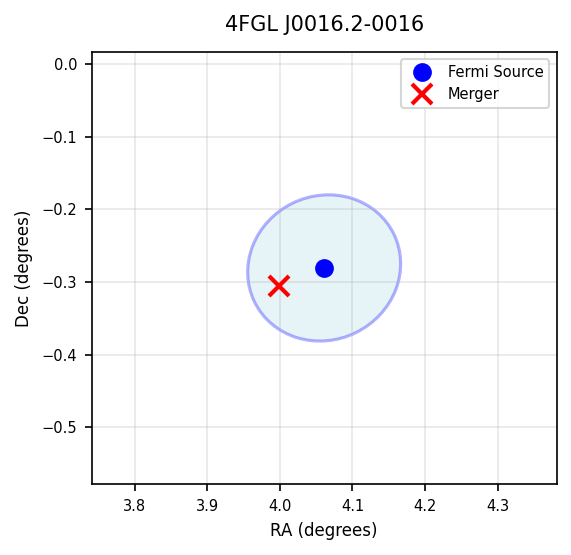}
        \caption*{Source 12}
    \end{subfigure}
    \begin{subfigure}[b]{0.18\textwidth}
        \includegraphics[width=\textwidth]{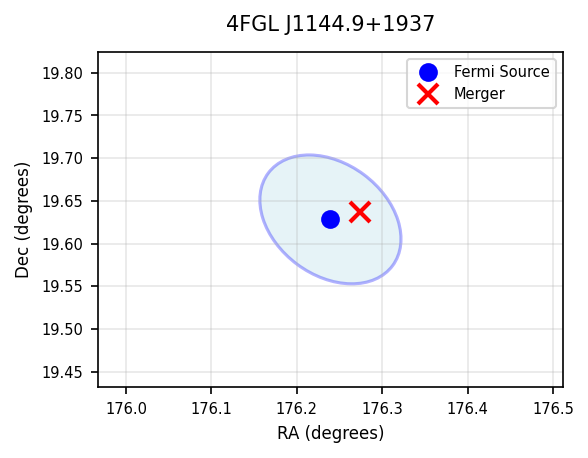}
        \caption*{Source 13}
    \end{subfigure}
    \begin{subfigure}[b]{0.18\textwidth}
        \includegraphics[width=\textwidth]{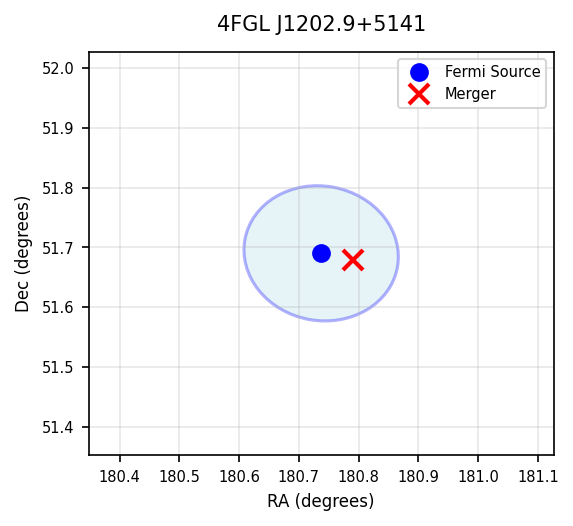}
        \caption*{Source 14}
    \end{subfigure}
    \begin{subfigure}[b]{0.18\textwidth}
        \includegraphics[width=\textwidth]{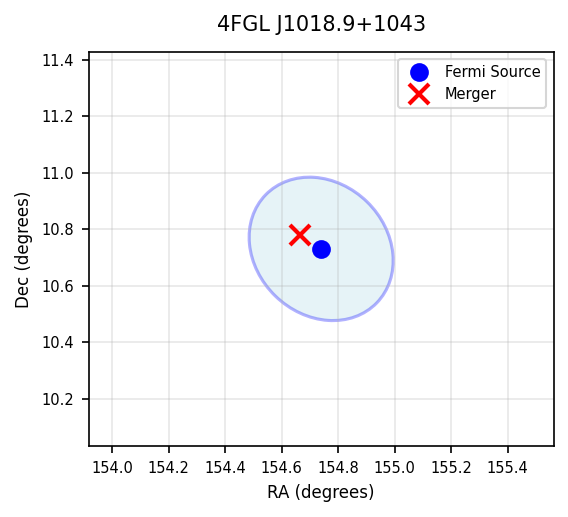}
        \caption*{Source 15}
    \end{subfigure}

    \vspace{0.2cm}

    \begin{subfigure}[b]{0.18\textwidth}
        \includegraphics[width=\textwidth]{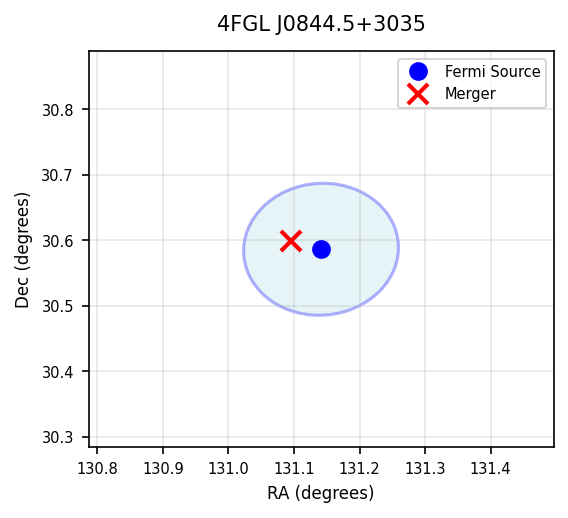}
        \caption*{Source 16}
    \end{subfigure}
    \begin{subfigure}[b]{0.18\textwidth}
        \includegraphics[width=\textwidth]{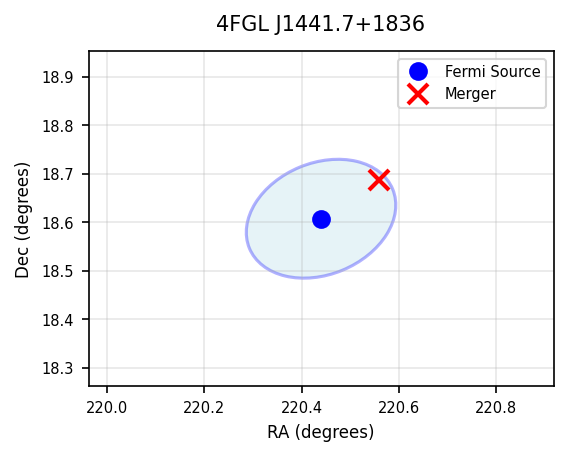}
        \caption*{Source 17}
    \end{subfigure}
    \begin{subfigure}[b]{0.18\textwidth}
        \includegraphics[width=\textwidth]{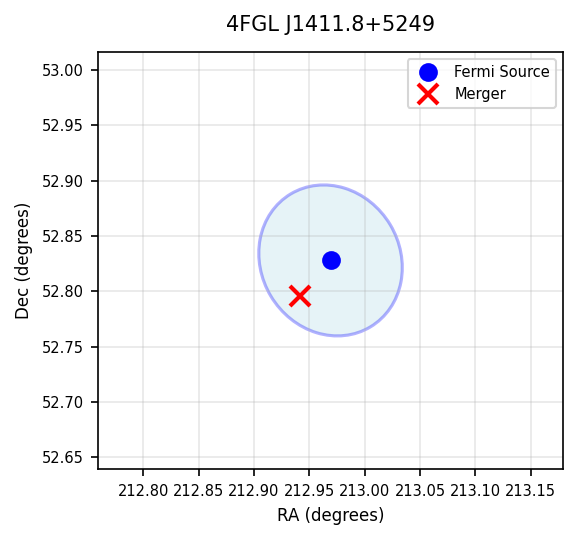}
        \caption*{Source 18}
    \end{subfigure}
    \begin{subfigure}[b]{0.18\textwidth}
        \includegraphics[width=\textwidth]{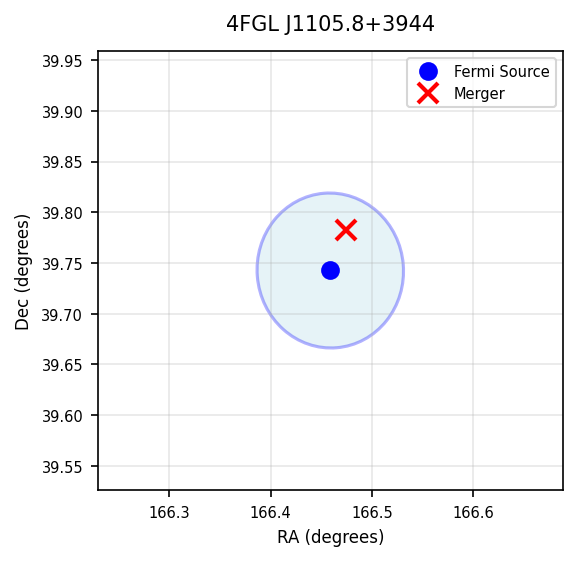}
        \caption*{Source 19}
    \end{subfigure}
    \begin{subfigure}[b]{0.18\textwidth}
        \includegraphics[width=\textwidth]{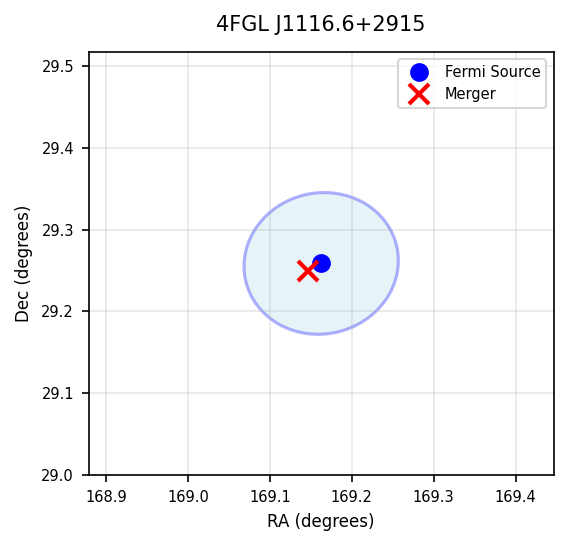}
        \caption*{Source 20}
    \end{subfigure}

    \vspace{0.2cm}

    \begin{subfigure}[b]{0.18\textwidth}
        \includegraphics[width=\textwidth]{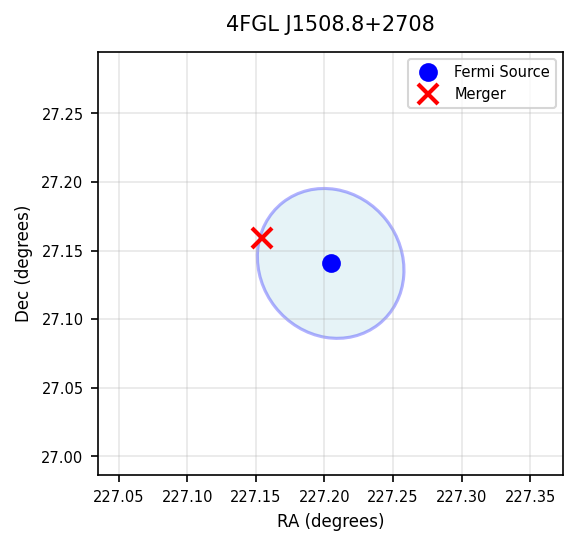}
        \caption*{Source 21}
    \end{subfigure}
    
   \caption{Merger positions located within the $4\sigma$ positional uncertainty ellipses of their associated Fermi sources. Each panel corresponds to a different matched Fermi source. The blue dot indicates the Fermi source position, while the red cross marks the galaxy merger with the lowest \emph{p}-value. The images are sorted in decreasing order of $p$-value, similar to Table~\ref{tab:secure_matches}.}
    \label{fig:4sigma_multipanel}
\end{figure*}

Overall, our analysis reveals that approximately 0.67\% (21 out of 3,166) of galaxy mergers exhibit statistically robust associations with gamma-ray sources. The predominance of blazar-type objects among these secure matches accounting for 62\% is consistent with their known dominance in the extragalactic gamma-ray sky. However, the presence of radio galaxies and a significant fraction of unassociated sources (24\%) suggests that merger-driven gamma-ray activity may originate from a broader range of physical mechanisms than previously recognized. To visualize these associations, Figure~\ref{fig:4sigma_multipanel} presents a multi-panel view of Fermi sources with at least one galaxy merger candidate within the corresponding $4\sigma$ positional uncertainty ellipse. In each panel, the ellipse denotes the Fermi positional uncertainty region, the blue dot marks the Fermi source position, and the red cross indicates the galaxy merger counterpart with the lowest \emph{p}-value. The plots are depicted in decreasing order of \emph{p}-value.  Moreover, in Table~\ref{tab:secure_matches} we show all the secure matches with their coordinates, SDSS $r$-mag, spectroscopic redshifts, flux densities, \emph{p}-values, angular separation and classification. Uncertainties in $r$-band magnitudes are from SDSS DR16 photometry; redshift uncertainties are from SIMBAD; flux errors are calculated from $r$-band magnitudes. To explore marginal cases, we extended the analysis to include matches with $p < 0.1$, identifying 24 potential associations. Of these, 21 are already included in the secure sample ($p < 0.05$). The remaining three additional sources are \textit{4FGL J1454.0+4927} (BCU), \textit{4FGL J1523.2+2818} (BCU), and \textit{4FGL J1223.3+1213} (BLL), with gamma-ray flux densities of $4.32 \times 10^{-36}$, $8.96 \times 10^{-36}$, and $4.79 \times 10^{-36}~\mathrm{erg\,cm^{-2}\,s^{-1}}$, and corresponding \emph{p}-values of 0.069, 0.071, and 0.076, respectively. These sources are spatially associated with the SDSS galaxies \textit{SDSS J145506.73+493614.7}, \textit{SDSS J152228.74+281402.9}, and \textit{SDSS J122256.10+121739.2}, which have spectroscopic redshifts of 0.0945, 0.0715, and 0.0905, respectively.


Furthermore, to assess the nature of the gamma-ray emission in the five unassociated Fermi sources in our sample, we examined their possible counterparts in the SDSS, the SIMBAD astronomical database~\citep{Wenger2000}, the ROSAT All-Sky Survey (RASS)~\citep{Voges1999}, and the Faint Images of the Radio Sky at Twenty Centimeters (FIRST) survey~\citep{Becker1994}. 
Three sources appear to be associated with galaxies showing no evidence of AGN activity, while the remaining two show signs of contamination by background AGN or blazars~\citep{Massaro2009,Dabrusco2019,Liu2011}.
4FGL~J0844.5+3035 is located 0.84 arcseconds from SDSS~J084422.83+303556.3, a galaxy at redshift $z = 0.074$ with no signs of AGN activity in its optical spectrum from SDSS DR16. The same galaxy is also listed as LEDA~1910153 in the Lyon-Meudon Extragalactic Database (LEDA), with a consistent redshift ($z = 0.07433$)~\citep{Zheng2020}. It is undetected in both the RASS and FIRST surveys, disfavoring a background AGN or blazar scenario. Similarly, 4FGL~J1112.0+2607 is offset by just 0.51 arcseconds from SDSS~J111204.44+260240.8, a galaxy at redshift $z = 0.06969$ in SDSS DR16, with no optical signs and no detections in the RASS or FIRST catalogs~\citep{Duarte2017}. Also, 4FGL~J1600.4+0407 is just 0.14 arcseconds from SDSS~J160049.74+041105.2, a galaxy at redshift $z = 0.08776$ with no AGN features in its SDSS spectrum and no detection in RASS or FIRST~\citep{Duarte2017}. These cases represent plausible candidates for merger driven gamma-ray emission.

In contrast, the other two sources show evidence of AGN associations. 4FGL~J1105.8+3944 lies only 0.15 arcseconds from SDSS~J110553.81+394656.9, a galaxy at redshift $z = 0.099$ with no AGN features in its SDSS DR16 optical spectrum~\citep{Toba2014,Kozie2020}. However, SIMBAD identifies a known BL Lac object, 87GB~110306.7+400253, at the same position~\citep{Massaro2009,Dabrusco2019}. 4FGL~J1435.4+3338 is found within 1.0 arcsecond of 2MASS~J14345466+3349199, which is classified in SIMBAD as an AGN with redshift $z = 0.05869$~\citep{Liu2011}. The nearest SDSS source, SDSS~J143454.21+334934.5, is a Seyfert 2 galaxy at $z = 0.05776$ based on its optical spectrum~\citep{Toba2014,Gaia2020}. Future dedicated multiwavelength follow-up, particularly with high sensitivity instruments such as eROSITA (X-ray) or the Very Large Array (VLA) would be valuable to further confirm the absence of AGN signatures and establish the nature of the emission.

To evaluate the possibility of angular clustering among the 21 matched gamma-ray--merger pairs, we performed a two-point angular separation analysis. Our results indicate that the mean angular separation among all unique pairs is $55.64^\circ$, with a median of $48.08^\circ$. The mean nearest-neighbor distance is $15.52^\circ$. Only 6 source pairs are within $10^\circ$, and just one pair lies within $5^\circ$. There are no pairs with separations less than $1^\circ$. 
These results suggest mild angular clustering, but the effect is not strongly concentrated, and the majority of sources are widely dispersed across the sky. We conclude that the observed clustering is weak and unlikely to bias our match probability estimates, which are computed individually based on local error ellipse geometry. This analysis has been included in the manuscript for completeness.

To estimate the expected number of false associations under a null hypothesis, we constructed a control sample by spatially shifting the sky coordinates of all 3,166 merger candidates by random angular offsets within $\pm 2^\circ$ in both Right Ascension and Declination. This procedure preserves the sky distribution but breaks any physical associations between the mergers and Fermi sources. The shifted catalog was then cross-matched against the Fermi 4FGL-DR4 catalog using the same 4$\sigma$ error ellipse matching criterion employed for the real sample. In this shifted control sample, we found zero matches. In contrast, the original unshifted sample yielded 21 matches with $p$-value $< 0.05$. This result supports the interpretation that the observed matches are not random coincidences but instead indicate a statistically significant excess.

The predominance of blazar-type classifications (FSRQ, BLL) among our matched sources aligns with recent theoretical work suggesting that gamma-ray emission in merging SMBH systems arises from jet interactions and magnetohydrodynamic instabilities triggered by gravitational perturbations from secondary black holes~\footnote{Foord, A. (2024). A High-Energy Perspective on Merging SMBHs. 11th International Fermi Symposium. \url{https://fermi.gsfc.nasa.gov/science/mtgs/symposia/eleventh/program/mon/Foord_AHighEnergyPerspectiveonMergingSMBHs.pdf}}.

\section{Conclusions}
\label{sec:conclusion}
We have presented the first dedicated cross-matching analysis between galaxy merger catalogs and the Fermi-LAT 4FGL-DR4 gamma-ray source catalog~\citep{Ballet2023}. From the first catalog, consisting of 70 galaxy pairs from the Canadian Network for Observational Cosmology Field Galaxy Redshift Survey~\citep{Patton2005}, we found no statistically significant associations within the 4$\sigma$ positional uncertainty threshold ($p < 0.05$). The second catalog comprises a high-confidence sample of 3,166 morphologically selected galaxy mergers derived from the initial catalog of 328,151 candidate mergers from~\citep{Ackermann2018}. Cross-matching with 4FGL-DR4~\citep{Ballet2023} under a conservative 4$\sigma$ positional criterion yielded 38 spatial coincidences, of which 21 show \emph{p}-values below 0.05. These associations are dominated by well-known gamma-ray source classes such as FSRQs, BL Lacs, and radio galaxies, lending support to the hypothesis that merger-driven processes can fuel AGN activity observable in gamma rays~\citep{Ellison2008,Ellison2011,Scudder2012,Satyapal2014,Gabor2016,Kashiyama2014}. Importantly, five secure associations correspond to unassociated Fermi sources, suggesting that merging systems may host previously unrecognized gamma-ray populations.

\section{Discussion and Outlook}
\label{sec:Discussion and Outlook}
While blazar-like AGN dominate the matched sample, the presence of radio galaxies, UIDs, and the broad range of observed gamma-ray flux densities point to a diversity of possible emission mechanisms. These may include merger-induced starbursts, large-scale shocks, or obscured AGN activity~\citep{Torres2004,Ackermann2012}, underscoring the importance of deeper multi-wavelength follow-up campaigns. The recurrent appearance of unassociated sources within the secure sample highlights the potential of galaxy merger catalogs to serve as physically motivated priors for gamma-ray source identification. 

As gamma-ray localization \rthis{improves} and forthcoming optical surveys such as the Vera Rubin LSST, Euclid, and DESI expand the census of galaxy mergers, the prospects for establishing robust associations \rthis{should improve} significantly. In addition, we plan to conduct a targeted search for specific merger candidates, such as UGC~12914/5, UGC~813/6, and VV~114, in a forthcoming work. Furthermore, future analyses could benefit from incorporating the future updated Fermi-LAT source catalogs as well as updated merger catalogs from ongoing and upcoming surveys, which will provide improved completeness and accuracy for cross-matching studies.

\begin{table}[htbp]
\centering
\caption{Complete list of 21 secure matches between Fermi 4FGL-DR4 sources and galaxy mergers, defined by $p$-value $<$ 0.05. 
The table is sorted in decreasing order of $p$-value. Uncertainties in $r$-band magnitudes are from SDSS DR16 photometry; redshift uncertainties are from SIMBAD; flux errors are calculated from $r$-band magnitudes.}
\resizebox{\textwidth}{!}{%
\begin{tabular}{lccccccccc}
\hline
\textbf{Fermi Source} & \textbf{Merger RA} & \textbf{Merger Dec} & \textbf{$r$-mag} & \textbf{z} & \textbf{Flux (erg/cm$^2$/s/Hz)} & \textbf{$p$-value} & \textbf{Separation} & \textbf{Class} \\
\hline
4FGL J1606.5+2717 & 241.6173 & 27.1373 & $17.119 \pm 0.007$ & $0.0458 \pm 0.00008$ & $5.15 \pm 0.033 \times 10^{-36}$ & 0.0480 & 3.70$\sigma$ & FSRQ \\
4FGL J1245.8+0232 & 191.5785 & 2.6659 & $17.394 \pm 0.006$ & $0.0811 \pm 0.00003$ & $4.00 \pm 0.022 \times 10^{-36}$ & 0.0344 & 3.31$\sigma$ & BLL \\
4FGL J0815.6+3641 & 123.8059 & 36.7766 & $16.851 \pm 0.007$ & $0.0420 \pm 0.00007$ & $6.60 \pm 0.043 \times 10^{-36}$ & 0.0252 & 3.77$\sigma$ & FSRQ \\
4FGL J1518.6+0614 & 229.5286 & 6.3552 & $16.892 \pm 0.008$ & $0.0795 \pm 0.00012$ & $6.35 \pm 0.047 \times 10^{-36}$ & 0.0274 & 2.68$\sigma$ & RDG \\
4FGL J1112.0+2607 & 168.0185 & 26.0447 & $17.046 \pm 0.005$ & $0.0696 \pm 0.00001$ & $5.51 \pm 0.025 \times 10^{-36}$ & 0.0182 & 2.64$\sigma$ & UID \\
4FGL J1435.4+3338 & 218.7278 & 33.8222 & $15.394 \pm 0.003$ & $0.0586 \pm 0.00001$ & $1.90 \pm 0.017\times 10^{-35}$ & 0.0171 & 3.00$\sigma$ & UID \\
4FGL J0813.4+2531 & 122.8163 & 25.1794 & $14.146 \pm 0.002$ & $0.0137 \pm 0.00001$ & $7.97 \pm 0.015 \times 10^{-35}$ & 0.0113 & 3.99$\sigma$ & FSRQ \\
4FGL J1008.0+0028 & 151.9620 & 0.5303 & $17.075 \pm 0.006$ & $0.0934 \pm 0.00003$ & $5.37 \pm 0.030 \times 10^{-36}$ & 0.0111 & 3.13$\sigma$ & BLL \\
4FGL J1145.5$-$0340 & 176.3724 & $-$3.6637 & $16.811 \pm 0.005$ & $0.0790 \pm 0.00001$ & $6.84 \pm 0.032 \times 10^{-36}$ & 0.0122 & 0.83$\sigma$ & BLL \\
4FGL J1600.4+0407 & 240.2073 & 4.1848 & $17.213 \pm 0.009$ & $0.0877 \pm 0.00001$ & $8.14 \pm 0.039 \times 10^{-36}$ & 0.0137 & 3.42$\sigma$ & UID \\
4FGL J1353.2+3740 & 208.3187 & 37.7076 & $16.675 \pm 0.005$ & $0.0630 \pm 0.001$ & $7.76 \pm 0.036 \times 10^{-36}$ & 0.0076 & 1.26$\sigma$ & BLL \\
4FGL J0016.2$-$0016 & 3.9988 & $-$0.3049 & $16.498 \pm 0.006$ & $0.0396 \pm 0.00004$ & $9.13 \pm 0.050 \times 10^{-36}$ & 0.0081 & 2.50$\sigma$ & FSRQ \\
4FGL J1144.9+1937 & 176.2744 & 19.6365 & $17.479 \pm 0.008$ & $0.0162 \pm 0.00004$ & $3.70 \pm 0.027 \times 10^{-36}$ & 0.0083 & 1.97$\sigma$ & RDG \\
4FGL J1202.9+5141 & 180.7901 & 51.6782 & $15.849 \pm 0.003$ & $0.0607 \pm 0.00001$ & $1.66 \pm 0.046 \times 10^{-35}$ & 0.0060 & 1.67$\sigma$ & BCU \\
4FGL J1018.9+1043 & 154.6662 & 10.7793 & $14.790 \pm 0.002$ & $0.0513 \pm 0.00001$ & $4.40 \pm 0.081 \times 10^{-35}$ & 0.0058 & 0.53$\sigma$ & BCU \\
4FGL J0844.5+3035 & 131.0952 & 30.5990 & $15.716 \pm 0.003$ & $0.0748 \pm 0.00001$ & $1.87 \pm 0.052 \times 10^{-35}$ & 0.0042 & 1.65$\sigma$ & UID \\
4FGL J1441.7+1836 & 220.5586 & 18.6907 & $15.233 \pm 0.003$ & $0.0710 \pm 0.00019$ & $2.92 \pm 0.081 \times 10^{-35}$ & 0.0035 & 3.91$\sigma$ & BCU \\
4FGL J1411.8+5249 & 212.9420 & 52.7960 & $16.248 \pm 0.004$ & $0.0766 \pm 0.00002$ & $1.15 \pm 0.042 \times 10^{-35}$ & 0.0028 & 2.64$\sigma$ & BLL \\
4FGL J1105.8+3944 & 166.4743 & 39.7825 & $15.958 \pm 0.004$ & $0.0991 \pm 0.00002$ & $1.50 \pm 0.055 \times 10^{-35}$ & 0.0026 & 2.26$\sigma$ & UID \\
4FGL J1116.6+2915 & 169.1460 & 29.2497 & $14.657 \pm 0.003$ & $0.0500 \pm 0.00001$ & $4.97 \pm 0.014 \times 10^{-35}$ & 0.0006 & 0.80$\sigma$ & RDG \\
4FGL J1508.8+2708 & 227.1547 & 27.1589 & $15.109 \pm 0.004$ & $0.0760 \pm 0.00002$ & $3.28 \pm 0.012 \times 10^{-35}$ & 0.0005 & 3.41$\sigma$ & BLL \\
\hline
\end{tabular}
}
\label{tab:secure_matches}
\end{table}

\section*{Acknowledgments}
SM thanks the Ministry of Education (MoE), Government of India, for their consistent support through the research fellowship, which has been instrumental in facilitating the successful completion of this work.
The authors also thank Lizhong Zhang for useful correspondence. We also thank the anonymous referee for very useful and constructive feedback on our manuscript.

\bibliographystyle{elsarticle-harv}
\bibliography{references}
\appendix
\section{SDSS Flux Conversion Methodology}
\label{appendix:sdss_flux_conversion}

SDSS magnitudes are reported in the {\tt asinh} magnitude system~\citep{Lupton1999}, which handle low S/N and negative fluxes more robustly than traditional Pogson magnitudes. The inverse asinh formula for converting $r$-band apparent magnitudes ($m_r$) to fluxes is:
\begin{equation}
    \frac{f_r}{f_0} = 2b \, \sinh\left(\frac{-2.5 \, m_r}{\ln(10)} - \ln(b)\right),
\end{equation}
where $f_0$ is the reference flux (1 maggy by definition), and $b = 1.2 \times 10^{-10}$ is the softening parameter for the $r$-band (see Table 21 in~\citep{Stoughton2002}). 

For completeness, the forward {\tt asinh} magnitude definition is:
\begin{equation}
    m = -\frac{2.5}{\ln(10)} \left[ \text{asinh}\left(\frac{f/f_0}{2b}\right) + \ln(b) \right].
\end{equation}

The resulting fluxes, in units of nanomaggies, were converted to Janskys using the SDSS relation:
\[
1\,\mathrm{nMgy} = 3.631 \times 10^{-6}\,\mathrm{Jy},
\]
and subsequently to cgs units via:
\[
1\,\mathrm{Jy} = 10^{-23}\,\mathrm{erg\,s^{-1}\,cm^{-2}\,Hz^{-1}}.
\]

\end{document}